\documentclass{JAIS}

\journal{JAIS-ID}
\vol{2023}

\received{31 September 2023}
\published{xx March 2024}

\def\be{\begin{equation}}
\def\ee{\end{equation}}
\def\bea{\begin{eqnarray}}
\def\eea{\end{eqnarray}}

\usepackage{caption}
\usepackage{comment}
\usepackage{lineno}
\usepackage{hyperref}
\usepackage{subcaption}

\articletype{Technical Report}
\begin{document}
%\title{Status and first results of the MURAVES campaign at Mt. Vesuvius}
%\title{Simulation Tools for the MURAVES Experiment}
\title{Simulation tools, first results and experimental status of the MURAVES experiment}

\author{Marwa Al Moussawi\auno{1}, 
Fabio Ambrosino\auno{2,3}, 
Antonio Anastasio\auno{2}, 
%Guglielmo Baccani\auno{4}, 
Samip Basnet\auno{1}, 
Lorenzo Bonechi\auno{4}, 
Massimo Bongi\auno{4,5}, 
Diletta Borselli\auno{4,6},
Alan Bross\auno{7}, 
Antonio Caputo\auno{8}, 
Roberto Ciaranfi\auno{4}, 
Luigi Cimmino\auno{2,3}, 
Vitaliano Ciulli\auno{4,5}, 
Raffaello D'Alessandro\auno{5}, 
Mariaelena D'Errico\auno{2,3},
Catalin Frosin\auno{4,5},
Andrea Giammanco\auno{1}, 
Flora Giudicepietro\auno{8}, 
Sandro Gonzi\auno{4,5}, 
Yanwen Hong\auno{9}, 
Giovanni Macedonio\auno{8}, 
Vincenzo Masone\auno{2}, 
%Nicola Mori\auno{5}, 
Massimo Orazi\auno{8}, 
Andrea Paccagnella\auno{4,5},
Rosario Peluso\auno{8}, 
Anna Pla-Dalmau\auno{7}, 
%Cesar Rendon Hinestroza\auno{8}, 
Amrutha Samalan\auno{9}, 
Giulio Saracino\auno{3}, 
Giovanni Scarpato\auno{8}, 
Paolo Strolin\auno{2,3}, 
Michael Tytgat\auno{9,10}, 
Enrico Vertechi\auno{8}, 
Lorenzo Viliani\auno{4}}
\address{$^1$Centre for Cosmology, Particle Physics and Phenomenology, Université catholique de Louvain, Belgium}
\address{$^2$National Institute of Nuclear Physics - Section of Naples, Italy}
\address{$^3$Department of Physics, University of Naples Federico II, Italy}
\address{$^4$National Institute of Nuclear Physics - Section of Florence, Italy}
\address{$^5$Department of Physics, University of Florence, Italy}
\address{$^6$Department of Physics and Geology, University of Perugia, Italy}
\address{$^7$Fermi National Accelerator Laboratory, USA}
\address{$^8$National Institute of Geophysics and Volcanology - Vesuvius Observatory, Italy}
\address{$^9$Department of Physics and Astronomy, Ghent University, Belgium}
\address{$^{10}$Department of Physics, Vrije Universiteit Brussel, Belgium
\\
\vspace{0.2cm}
Corresponding authors: Andrea Giammanco, Yanwen Hong\\
Email address: andrea.giammanco@uclouvain.be, yanwen.hong@cern.ch
}

\begin{abstract}
The MUon RAdiography of VESuvius (MURAVES) project aims at the study of Mt. Vesuvius, an active and hazardous volcano near Naples, Italy, with the use of muons freely and abundantly produced by cosmic rays. In particular, the MURAVES experiment intends to perform muographic imaging of the internal structure of the summit of Mt. Vesuvius. The challenging measurement of the rock density distribution in its summit by muography, in conjunction with data from other geophysical techniques, can help model possible eruption dynamics. The MURAVES apparatus consists of an array of three independent and identical muon trackers, with a total sensitive area of 3 square meters. In each tracker, a sequence of 4 XY tracking planes made of plastic scintillators is complemented by a 60 cm thick lead wall inserted between the two downstream planes to improve rejection of background from low energy muons. 
The apparatus is currently acquiring data. 
This paper presents preliminary results from the analysis of the first data samples acquired with trackers pointing towards Mt. Vesuvius, including the first relative measurement of the density projection of two flanks of the volcano at three different altitudes; we also present the workflow of the simulation chain of the MURAVES experiment and its ongoing developments.
\end{abstract}

\maketitle


\section{Introduction}

%{\bf Must be checked/expanded by Giovanni / Flora}

Mount Vesuvius is an active strato-volcano near Naples (Italy). 
Understanding its composition is very important for volcanology and civil protection. 
Famous for the eruption that buried Pompeii and Herculaneum in 79 C.E., at the present time it is considered one the most dangerous volcanoes in the world, as $>0.5$ million people reside in its surrounding area, which is at high risk of pyroclastic fallout in case of a Sub-Plinian eruption~\cite{BarberiRisk}. 
Vesuvius has a complex geological history, and underwent its latest major structural modification during the eruption of 1944. 
Gravimetry and seismic tomography have given discordant results about the deep structure of Mt.Vesuvius~\cite{cella2007shallow,de2006somma}. 

%{\bf ... state of the art in knowledge about Vesuvius by other techniques (Giovanni / Flora)...}

Muon radiography, or "muography" for short, is a subsurface remote-sensing technique based on the absorption of muons (elementary particles with the same quantum numbers as the electrons, but 200 times heavier) when passing through matter. 
In this technique, the attenuation of the cosmic muon flux is exploited to measure differences in average density, in a way that is conceptually similar to conventional (X-ray) radiography. 
The large penetration power of the muons (which lose roughly 200~MeV per meter of water equivalent) and their broad energy spectrum, which extends to O(TeV), make muography a promising method for the imaging of gigantic objects. 
A pioneering role in the imaging of mountains by muography has been played by Nagamine’s team in the 90's in Japan~\cite{nagamine1995method}. Since the beginning, this research was motivated by future applications to volcanoes, which followed in the next decade, and in 2009 muography was used for the first time, during an unrest of Mount Asama in Japan, to correlate the density map evolution of the volcano with its eruption sequence~\cite{tanaka2009detecting}. Nowadays, the list of volcanoes already actively studied by muography includes Asama~\cite{tanaka2019japanese}, Satsuma-Iwojima~\cite{tanaka2014radiographic} and Sakurajima~\cite{olah2019plug} in Japan, Vesuvius~\cite{ambrosino2014mu}, Etna~\cite{presti2018mev} and Stromboli~\cite{tioukov2019first} in Italy, as well as Puy de D\^ome~\cite{carloganu2013towards,ambrosino2015joint} and La Soufri\`ere de Guadeloupe~\cite{le2019abrupt} in France. 
Several Colombian volcanoes are soon going to join this list~\cite{guerrero2019design,vesga2017muon}. 
For recent reviews of several volcanology applications of muography, we refer the reader to \cite{MuographyBook}. Applications of muography to other fields have also been reviewed e.g. in \cite{Bonechi:2019ckl,IAEA2022}.

A pilot study of the prospects of muography on Mt. Vesuvius was published in 2014 by the MU-RAY project~\cite{ambrosino2014mu}, based on approximately one month of data. 
The MU-RAY muon telescope consisted of three layers of scintillator bars coupled to silicon photomultipliers (SiPM), each layer having a $1~m^2$ surface and composed of back-to-back orthogonal planes providing $x-y$ hit position. 
In addition to providing the very first 2D muography of Mt. Vesuvius, this first campaign gave operational experience and hinted at a serious pollution by background tracks that overwhelmed the signal in the image regions corresponding to the thickest parts of the volcanic cone. 
A crucial step in understanding the composition of the background was a joint data-taking campaign with the TOMUVOL collaboration~\cite{ambrosino2015joint}. The muon telescopes from the two projects, based on different technologies (resistive plate chambers in the case of TOMUVOL~\cite{carloganu2013towards}) and with different strategies for the reduction of backgrounds, took data simultaneously at the Puy de D\^ome, finding results consistent with each other and with the aforementioned pilot study of Mt. Vesuvius. This allowed to exclude large contamination from combinatorial background, to which TOMUVOL was largely insensitive by design, and from fake muons (such as electrons, positrons or protons) which MU-RAY rejected by a 3~cm steel plate, to induce showering of the fake muons, amounting to an effective threshold of 70~MeV in muon momentum. 
This led to the conclusion that this background was dominated by "soft" (i.e. low-energy) muons, undergoing large scattering in the slopes of the volcano or even back-scattering from the ground surrounding the muon telescopes; as their real trajectories are practically uncorrelated with the trajectories presumed in the high-level reconstruction, this results in an irreducible blurring of the muography images~\cite{gomez2017forward}. 
This hypothesis was then confirmed by Monte Carlo studies with PUMAS~\cite{PUMAS}, and the rejection of the sub-GeV component of the muon spectrum became a main consideration in the design of the MUon RAdiography of VESuvius (MURAVES) experiment~\cite{d2019volcanoes,saracino2017muraves}, whose status is reported in this document. 

This article is structured as follows: 
Section~\ref{sec:muraves} describes the MURAVES set-up that is currently taking data on Mt. Vesuvius; 
Section~\ref{sec:first-results} summarizes the earliest public results based on preliminary data; 
Section~\ref{sec:simulation} presents the current status of the work towards and end-to-end Monte Carlo simulation chain of the experiment, focusing on what is new since our previous reports on this specific subject~\cite{moussawi2022simulations,samalan2022end}. 
We conclude in Section~\ref{sec:conclusion} with the lessons learnt and some directions for improvement that have been identified.

\section{The MURAVES experiment}
\label{sec:muraves}

\begin{figure}
    \centering
    \includegraphics[height=0.35\linewidth]{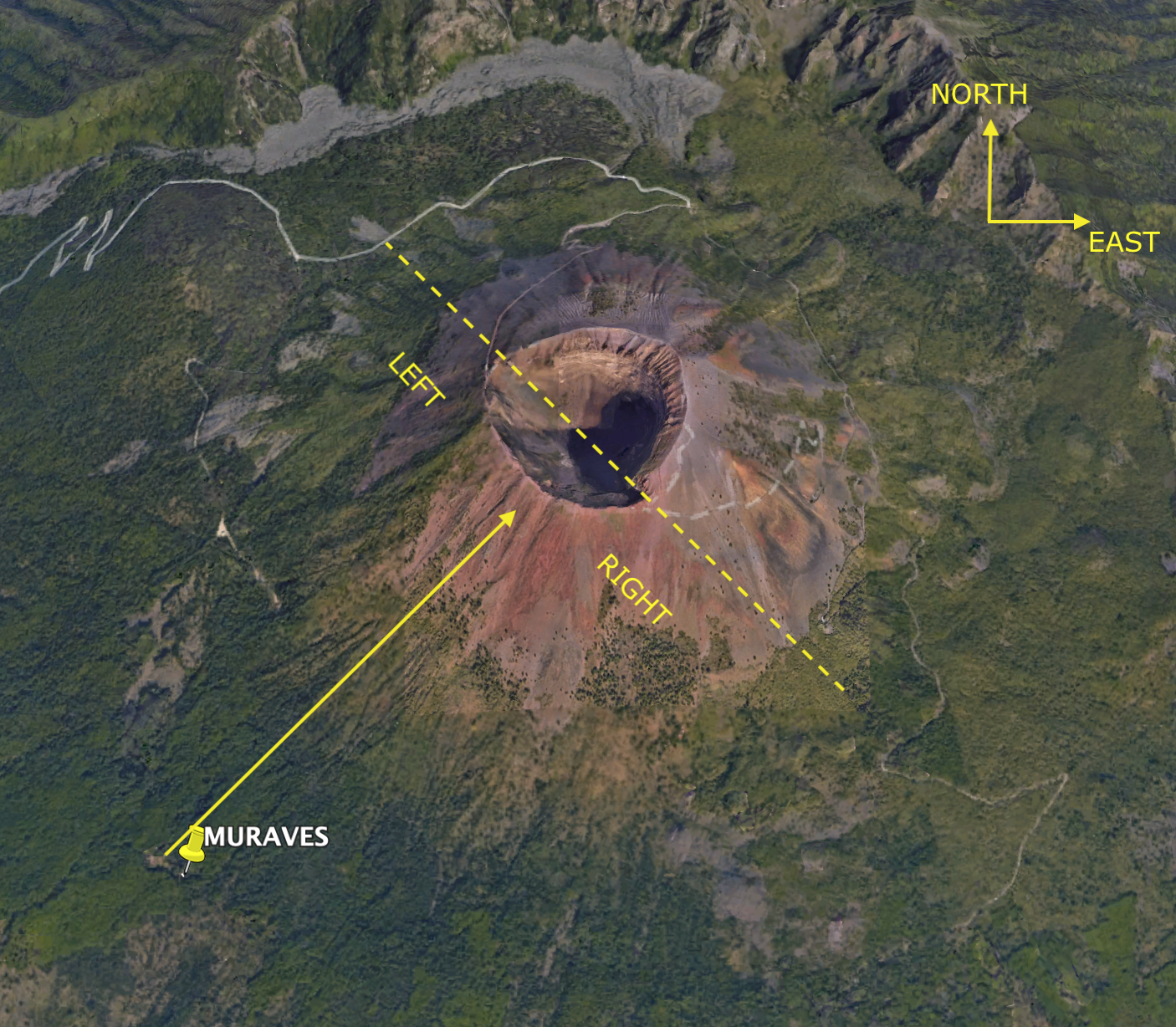}
    \includegraphics[height=0.35\linewidth]{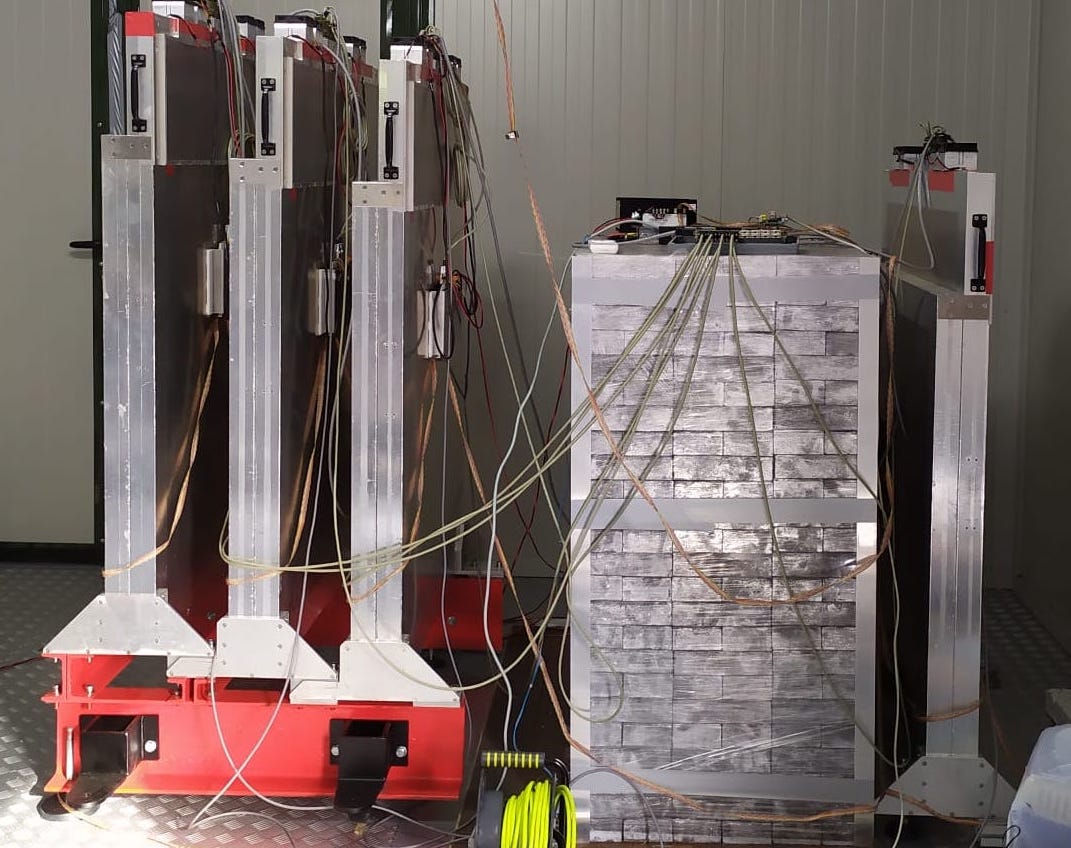}
    \caption{Left: location of the MURAVES site with respect to Mt. Vesuvius. Right: one of the MURAVES muon telescopes. Reproduced from Ref.~\cite{d2022muraves}.}
    \label{fig:muraves}
\end{figure}

The MURAVES experiment consists of three identical muon telescopes, labeled NERO, ROSSO and BLU, taking data simultaneously. The three telescopes are hosted in a solar-powered container, located on the South-West flank of the volcano at 600~m a.s.l.~\footnote{The summit of Mt. Vesuvius reaches 1281~m a.s.l.} and at a distance of 1500~m from the summit. 
The site, indicated in Fig.~\ref{fig:muraves} (left), has been chosen based on two main criteria: accessibility and signal-to-noise ratio, the latter assessed with PUMAS~\cite{PUMAS}, where the signal is defined as muons detected in the same angular bin where they are generated, while the background is defined as muons detected in a different bin.

As shown in Fig.~\ref{fig:muraves} (right), each muon telescope is composed of four layers (one more than the MU-RAY telescope), staggered in such a way that the acceptance of the telescope includes the Great Cone of Mt. Vesuvius,
with a 60~cm thick lead wall between the third and the fourth layer to act as a passive momentum filter, corresponding to a cut-off of $\approx 900$~MeV for muons impinging orthogonally. 

Each layer has the same size and layout as in the precursor project MU-RAY~\cite{ambrosino2014mu}, with $x-y$ spatial information coming from two orthogonally oriented planes of 64 plastic scintillator bars, whose light signal is collected by wavelength-shifting fibres and read-out by SiPMs. 
The scintillator bars have an isosceles triangle-shaped section, with 3.3~cm basis and 1.7~cm height. This shape ensures that each muon passes through at least two adjacent bars, hence its hit position can be estimated by the average of the bar centre positions weighted by the energy deposited in the bars, improving the spatial resolution. 
The SiPMs and the read-out of MURAVES have been upgraded with respect to MU-RAY.
As the SiPM response is highly sensitive to ambient temperature, a temperature control system based on Peltier cells plays an important role in MURAVES.

The container has four slots to host the three telescopes, each slot having an unmovable lead wall. All telescopes can be easily disassembled and reassembled in a different slot. 
Three of the slots are meant for the ``Vesuvius runs'' of the telescopes, while the fourth is for ``free-sky runs'', i.e. with no object in the field of view of the telescope. The practical difference is in the position of the lead wall, that in the former case allows for three detector layers to be on the Vesuvius side, and in the latter case the other way around. 
%{\bf FIXME: insert figure}
During the lifetime of the experiment, each telescope is planned to alternate Vesuvius and free-sky runs, as the latter are important as control data to verify the status of the detector and also for the reference flux measurements that are used in the transmission method to extract a 2D density projection (or opacity) map. 

Given the differential muon flux $I(E; \alpha$, $\phi)$ and a run duration $\Delta T$, the observed muon flux through the target can be expressed as follows:
\begin{equation} \label{eq:flux}
N_{\mu}(\alpha,\phi)=\Delta T \cdot \epsilon (E;\alpha,\phi)\cdot A(\alpha,\phi)\cdot \int^{\infty}_{E_{min}(X)} I(E;\alpha,\phi)dE \; ,
\end{equation}
where $\alpha$ is the elevation angle (related to the zenith angle $\theta$ by $\alpha = (\pi/2) - \theta$) and $\phi$ the horizontal angle of arrival of the muon, $\epsilon$ is a global detector efficiency that in principle depends on both direction ($\alpha$ , $\phi$) and $E$. 
Finally, $A$ is an acceptance factor that only depends on geometry and orientation of the detectors. 
The observed number of muons depends on the opacity ($X$) i.e. the projection of the density ($\rho$) along the line of sight:
\begin{equation}
X = \int^{exit}_{entry} \rho\; dx \; ,
\end{equation}
through the integration extreme in equation~\ref{eq:flux}:
\begin{equation}
E_{min}=E^{rock}_{min}(X) + E_{det}^{min} \; ,
\end{equation}
where $E^{rock}_{min}(X)$ is the minimum energy a muon needs to survive the opacity $X$, so it is also function of $(\alpha,\phi)$, while $E_{det}^{min}$ is the energy necessary to be detected. 
The MURAVES data analysis is based on muons that give hits in all four layers of a telescope, which implies that $E_{det}^{min}$ is dictated at first order by the lead wall~\footnote{Note, however, that the trigger logic is based on the presence of hits in the first three layers, which allows the possibility in the offline analysis stage to study the effect of the lead wall e.g. on the displacement of hits in the fourth layer~\cite{moussawi2022simulations}.}. 
Opacity depends on the density and on the thickness of the material traversed. As we are interested in the former, the latter is calculated for each $(\alpha,\phi)$ direction using a detailed Detector Terrain Model (DTM) of the volcano~\cite{samalan2022end}, based on data from Ref.~\cite{vilardo2013morphometry}, which has a horizontal resolution of 5~m and an altitude resolution of 1~m. %{\bf Giovanni, please confirm - I took these resolutions from Mariaelena's thesis}

In order to minimize the dependence on several modeling aspects, the opacity map is extracted in practice from the transmission, $T(\alpha,\phi)$, defined as the ratio between the measured flux in eq. (\ref{eq:flux}) and a reference flux measured with the free-sky runs:
\begin{equation}
T(\alpha,\phi)=\frac{N_{v}(\alpha,\phi)\Delta T_{fs}}{N_{fs}(\alpha,\phi)\Delta T_{v}} \; ,
\label{eq:transmission}
\end{equation}
where $N_{v}$ and $N_{fs}$ indicate the muon counts in a $(\alpha,\phi)$ bin in Vesuvius ($v$) and free-sky ($fs$) runs, respectively, while $\Delta T_{v}$ and $\Delta T_{fs}$ are the respective run durations. 
Using data from the same telescope and the same operating working point (see Sec.~\ref{sec:datasets}) in Vesuvius and free-sky runs, geometrical factors and trigger efficiency are in good approximation equal for both datasets, thus independent from acceptance and efficiencies. Transmission measurements from different telescopes or different working points can be safely combined under this assumption, while combining at the level of the raw counts of muons would introduce biases that are difficult to estimate reliably. 
\section{{Simulation tools}}
\label{sec:simulation}
Muography applied to volcanology poses significant challenges due to the substantial rock thicknesses found in volcanoes, which can extend up to several kilometers. To effectively measure the muon flux traversing the geometry, it is upon the presence of high-energy muons, as demonstrated in equation (\ref{eq:flux}). However, such high-energy muons are characterized by a low incidence rate.
Hence, precise predictions relying on rigorous simulation techniques become necessary. This chapter introduces the simulation tools employed within the MURAVES experiment. Three cosmic muon generators, namely CORSIKA \cite{CORSIKA}, CRY \cite{CRY}, and EcoMug \cite{EcoMug}, are studied and compared. The simulation of detector responses is carried out within the GEANT4 framework \cite{GEANT4}, and the whole simulated data processing chain is emulated. Notably, for the first time we compare two distinct transport engines, namely MUSIC \cite{MUSIC} and PUMAS \cite{PUMAS}, to simulate muon transport through the Mount Vesuvius, benchmarking them againt GEANT4.

\subsection{Cosmic Muon Generators}
Cosmic muon generation is undertaken across the CORSIKA, CRY, and EcoMug generators. Table \ref{t:muongenerator} summarizes their main characteristics and distinctions among these three generators. EcoMug and CRY adopt parametric simulation methodologies, grounded in specific experimental or simulated data, whereas CORSIKA perform a comprehensive, step-by-step evolution of cosmic showers, offering a range of low- and high-energy hadronic interaction models and a more realistic representation of multi-muon events, which is particularly relevant to our investigation.
However, modelling accuracy is not a crucial factor in volcanology applications of muography given the currently low statistical precision achievable, therefore CRY has been retained as the primary generator of choice due to its rapid response time and seamless integration with GEANT4 for subsequent stages of the simulation chain. The other two generators, EcoMug and CORSIKA, are maintained for the purpose of estimating systematic uncertainties. In our study, broadly speaking, all three generators exhibit a satisfactory degree of consistency in their results.

\begin{table}[h!]
\tbl{Assessment of three cosmic muon generators.\label{t:muongenerator}}{
\begin{tabular}{c||p{4cm}|p{4cm}|p{4cm}}
\hline
Generator &  EcoMug & CRY & CORSIKA \\
\hline\hline
Principle & parametric, one particle per event & parametric, few particles per event & full cosmic shower\\
\hline
Modeling & configurable; default based on ADAMO experiment~\cite{adamo} & fixed; based on MCNPX simulation~\cite{MCNPX} & various low- and high-energy hadronic interaction models provided \\
\hline
Generation Surface & flat, cylindrical and hemispherical & flat & flat \\
\hline
Speed ($10^{5}$ muons) & O (sec) & O (min) & O (hour) \\
\hline
Integration with GEANT4 & easy & easy & complex \\
\hline
\end{tabular}}
\end{table}

\subsection{Detector Simulation and Data Processing}
\begin{figure}[h]
\centering
\includegraphics[width=10cm]{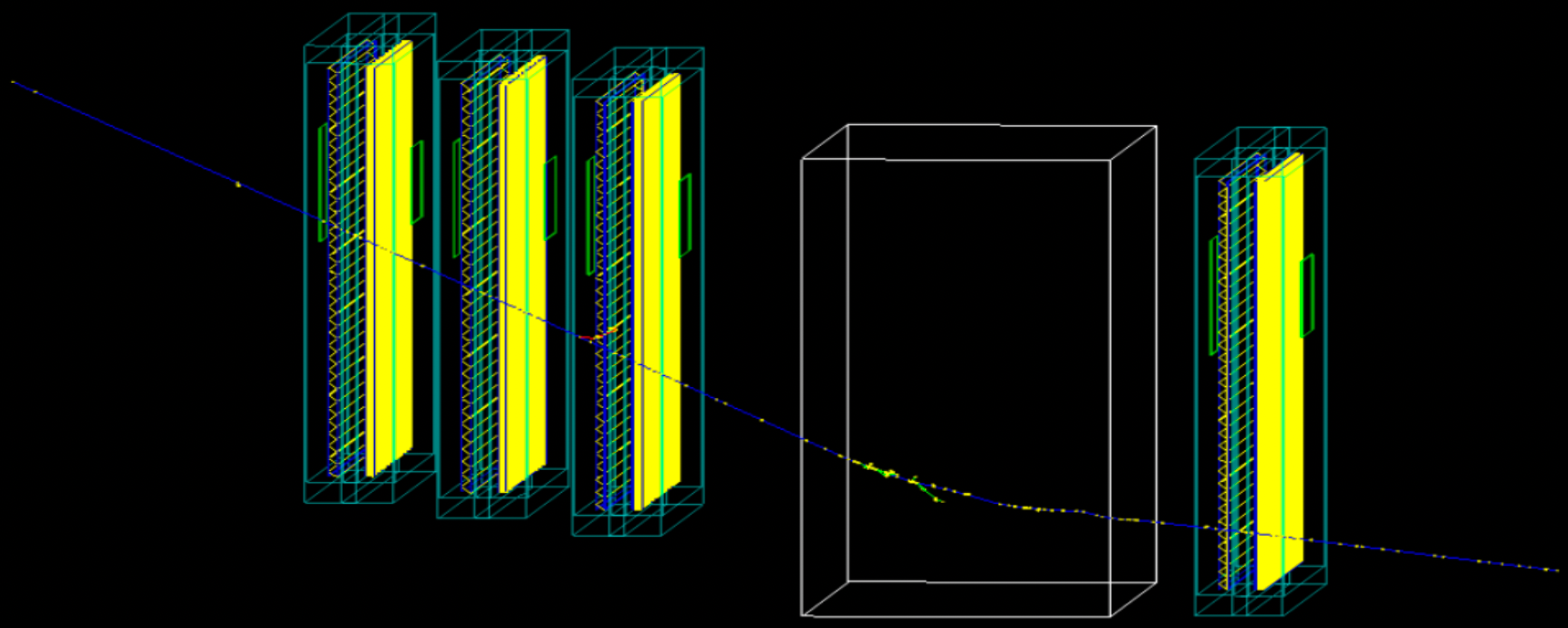}
\caption{GEANT4 model of a MURAVES telescope, traversed by a 1 GeV muon.}
\label{fig:detectorSimulation}
\end{figure}

The geometric configuration of the detector system has been characterized using GEANT4. As explained in section \ref{sec:muraves}, each muon telescope consists of three upstream stations, an 60 m intervening lead block, and the downstream station, distributed over $\approx 2$ m. Each station consists of a pair of orthogonal planes, where each plane is composed of 64 triangular scintillator bars. Scintillator light is collected via optical fibers and later read out by SiPM. 
During the course of the detector simulation,  a dataset of simulated ``hits'' is obtained, where each hit is defined by the three-dimensional spatial coordinates and the time of an interaction event, the energy deposited by the particle at that event, and the identity of that particle. Figure \ref{fig:detectorSimulation} is an instructive visualization depicting the trajectory of a 1 GeV muon as it traverses through a MURAVES hodoscope, displaying the trail of energy deposited.

\begin{figure}[h]
\centering
\includegraphics[width=10cm]{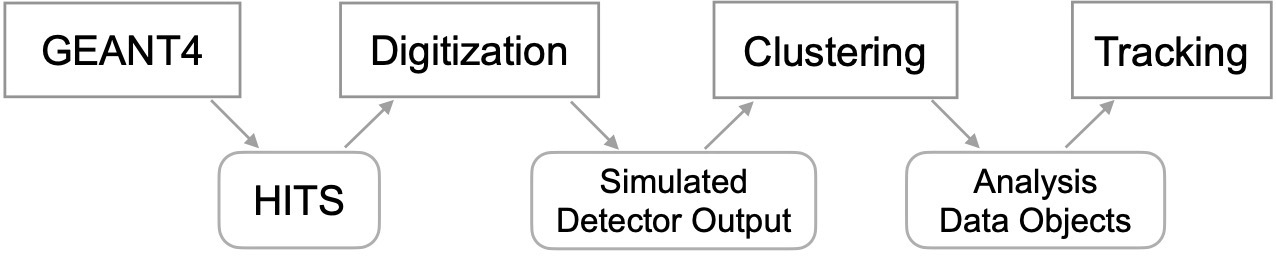}
\caption{Simulation data processing workflow for the MURAVES experiment. The top row indicates the algorithmic steps and the bottom row the data generated by each step and used as input by the next step.}
\label{fig:chain}
\end{figure}

Subsequently, within the simulated data processing workflow as indicated in Fig. \ref{fig:chain}, advanced techniques for digitization, clustering, and tracking are developed and performed. The digitization phase is responsible for the precise quantization of spatial positions and energy deposition events, resulting in simulated detector output data, encoded in exactly the same format as real raw data. Accordingly, the clustering step identifies and groups together adjacent detector elements whose signals exceed a predefined threshold. The output of the clustering process yields a set of analysis data objects, encompassing critical information such as the total energy deposited by the particle and its corresponding spatial coordinates, the latter estimated with the barycenter method i.e. through the weighted average of the center of the strip positions, where the weights are given by the signal amplitudes at each strip. Following the clustering phase, tracking methods are systematically applied to elucidate the trajectories and kinematic properties of the particles traversing the detector system, facilitating a comprehensive analysis of their interactions. 
Currently, tracks are fitted in the $xy$ and $xz$ planes independently; a novel tracking methodology is presently under development, which considers the sum the residuals resulting from the performed track fits on both planes. 
It is important to emphasize that, from the clustering step onward, the very same algorithms are applied in Monte Carlo and in real data.

\subsection{Muon Transport Through the Mt. Vesuvius}
The MUSIC~\cite{MUSIC} and PUMAS~\cite{PUMAS} libraries share a common objective, as transport engines for propagating muons through large thickness of rock or water. In the case of PUMAS, these transport capabilities extend to tau leptons as well. 
Both MC programs are designed such to minimize the time spent in the simulation of large datasets, which would be unfeasible by GEANT4, but they achieve that through very different strategies.

MUSIC is a parametric simulation. It only considers the main electromagnetic interactions causing energy loss, such as ionization, bremsstrahlung, electron-positron pair production, and muon-nucleus inelastic scattering. 
Its computational efficiency is enhanced by pre-computed and averaged muon interaction cross-sections for specified elements within materials. 
MUSIC is a mixed (class II) Monte-Carlo algorithm, i.e. it incorporates both soft and hard collisions of the muon in matter. 
An important tunable parameter is the threshold ($\nu_{thr}$) on the fraction of energy ($\nu$) expected to be lost by the muon. If $\nu < \nu_{thr}$, all collisions are considered as soft and their contribution to energy loss is approximated in a point and calculated with a continuous approximation, also known as the continuously slowing down approximation (CSDA), wherein energy losses follow deterministic principles. In this mode, the energy loss at each point along the track is assumed to be equal to the stopping power, disregarding any fluctuations in energy loss. For $\nu > \nu_{thr}$, instead, hard collisions are simulated stochastically. 
This is also called a Straggled mode. 
A recommended value for the threshold is $\nu_{thr} \approx 10^{-3}$.

Although PUMAS can be used as a traditional ``forward'' Monte Carlo (like GEANT4 and MUSIC), it also enables a comprehensive simulation process referred to as Backward Monte Carlo (BMC).
The BMC methodology involves the reversal of conventional practice utilized to generate a final state from an initial state, together with associated random variates; this reversal enables the expression of the initial state as function of the final state. This allows to save orders of magnitude in computation time by only simulating the muons that are usable in data analysis, ignoring those outside of acceptance. 
In addition, the PUMAS library offers distinct Monte Carlo transport modes. One of the modes available is the CSDA. Another mode is the Straggled mode, which employs a mixed MC algorithm similarly to the one employed in MUSIC.

\begin{figure}[h!]
     \centering
     \begin{subfigure}[h!]{0.45\textwidth}
         \centering
         \includegraphics[width=0.9\textwidth]{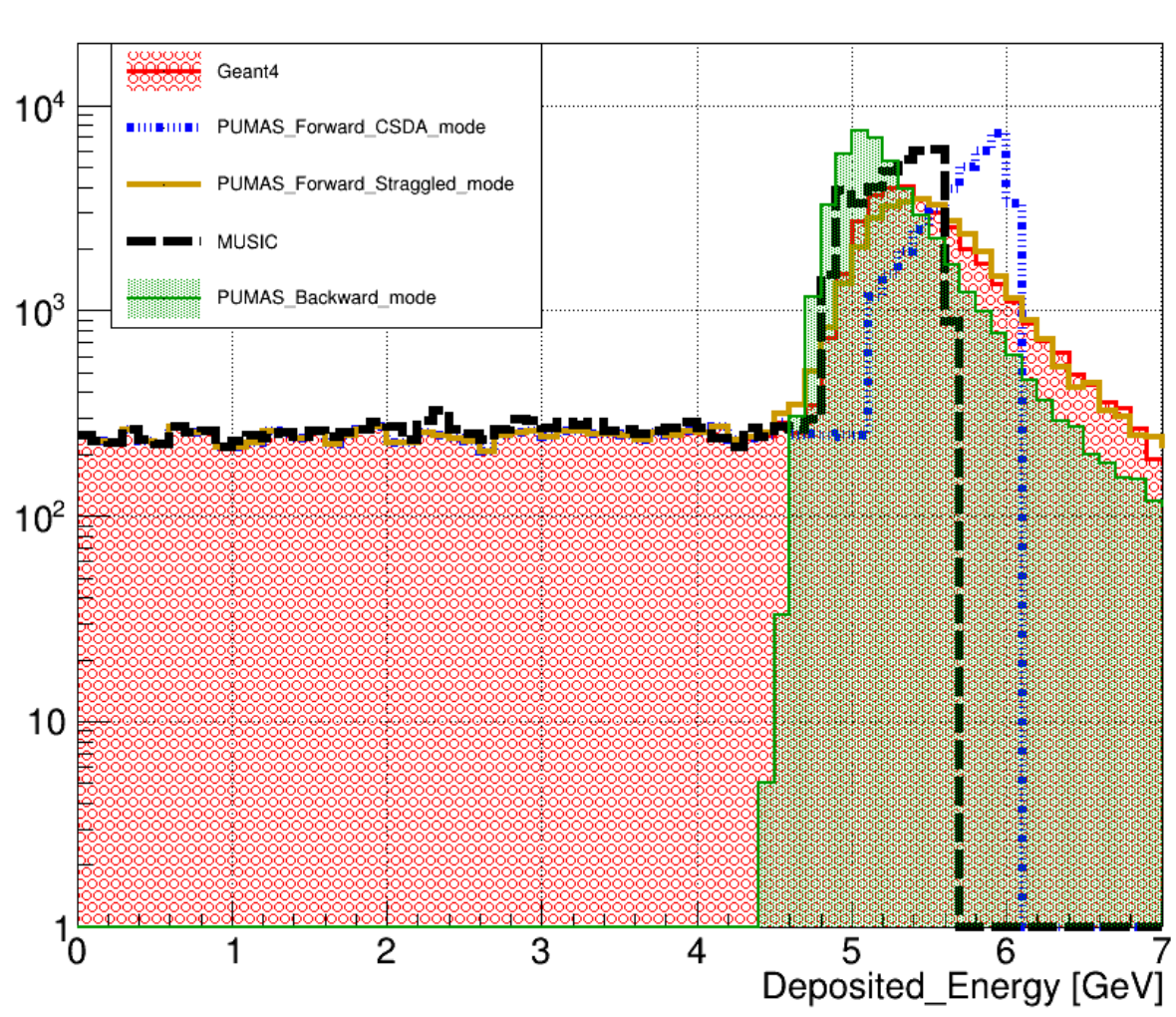}
         \caption{The final energy range of MUSIC output, but not the final energy distribution, is given to PUMAS backward mode.}
         \label{fig:comparisonPUMAS}
     \end{subfigure}
     \hfill
     \begin{subfigure}[h!]{0.45\textwidth}
         \centering
         \includegraphics[width=0.9\textwidth]{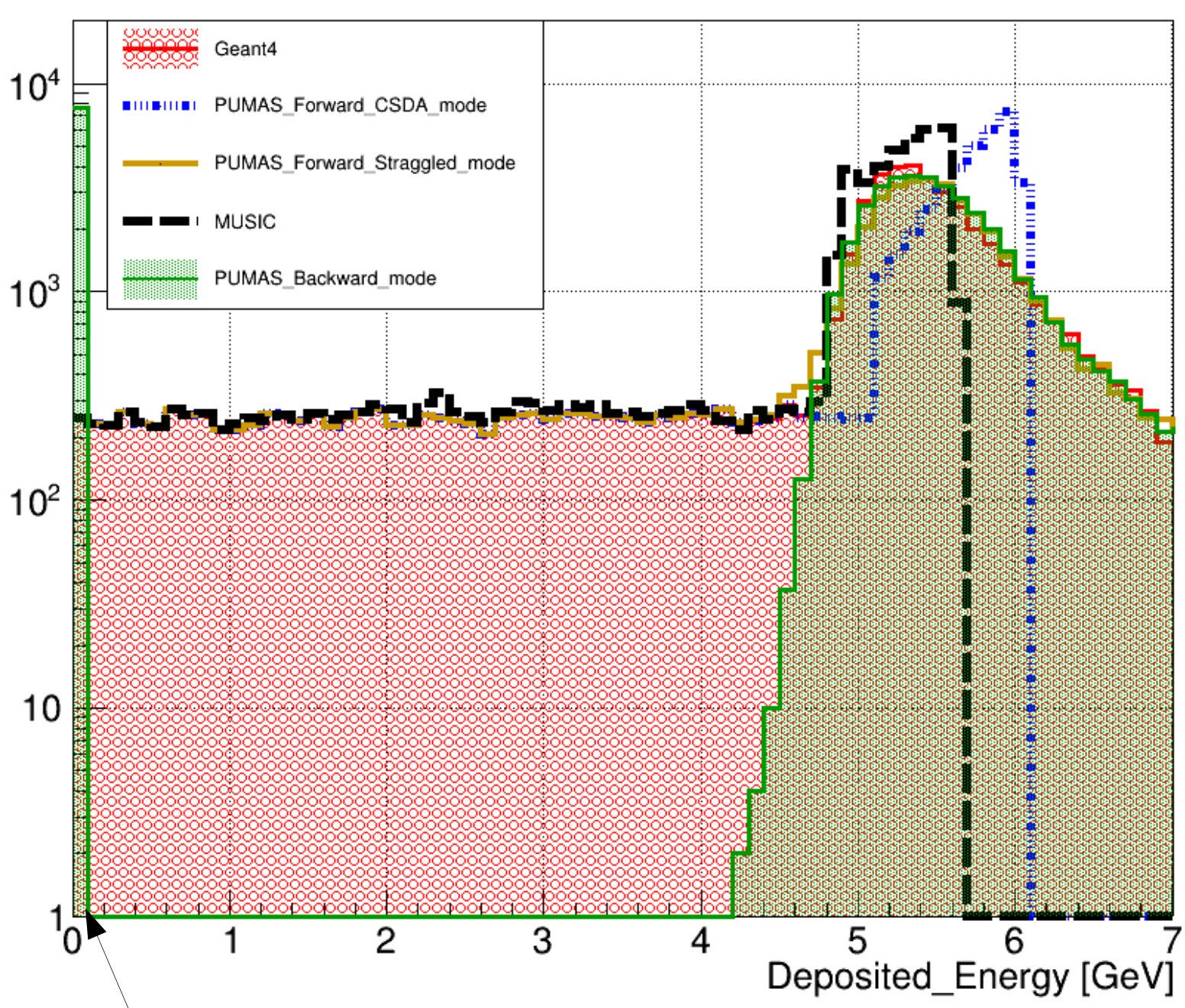}
         \caption{The exact final energy distribution of MUSIC output is given to PUMAS backward mode.}
         \label{fig:comparisonMUSIC}
     \end{subfigure}
        \caption{Comparison of energy loss distribution in GEANT4, MUSIC and different modes in PUMAS, with initial energy 0 - 20 GeV muon in a standard 10 m thick rock (2.65 $g/cm^{3}$).}
        \label{fig:comparison}
\end{figure}

\subsubsection*{Simplified Case Study}

To gain deeper insights into the functionalities of both transport engines, studies were conducted with a focus on simulation of muon energy loss in a standard rock. 
A simplified case study has been defined, and GEANT4 was used as benchmark to compare with MUSIC and with PUMAS in backward and forward modes, and in the latter case both CSDA and Straggled simulations have been executed. 
Muons have been generated with a uniform spectrum of initial energy ranging from 0 to 20 $GeV$, and their passage through a uniform slab of standard rock, characterized by a thickness of 10 $m$ and a density of 2.65 $g/cm^{3}$, has been simulated. 
The outcomes are compared in Fig.~\ref{fig:comparison}.

PUMAS, in backward mode, needs an energy distribution as input for reweighting, which needs therefore to be computed beforehand, e.g. by running GEANT4 or MUSIC, or with analytical approximate formulas. 
In order to illuminate the behavior of the backward mode, the following workflow was considered:
\begin{enumerate}
\item The final energy distribution, i.e. after passing through the rock, was simulated using MUSIC.
\item In one case (figure~\ref{fig:comparisonPUMAS}), only the range of the final energy (in our case 0 - 14.462~GeV) was used but the distribution was assumed to be constant; in another trial (figure~\ref{fig:comparisonMUSIC}) the exact distribution of the final energy estimated from MUSIC was used as input.
%\item The outcomes of the PUMAS backward mode were analysed and compared with all other MC options, as shown in Fig.~\ref{fig:comparison}.
\end{enumerate}

As shown in Fig.~\ref{fig:comparison}, the peak positions of the energy loss in different simulators are all relatively close, with an expectation value of 5.3~GeV. 
PUMAS in forward straggled mode can be seen to follow very closely the behaviour of GEANT4 across the full spectrum.
MUSIC, as well as PUMAS in forward CSDA mode, feature a sharp cut after the peak position, which can be explained by the fact that the additional pathlength due to hard scattering events is ignored in this approximation. 
While their qualitative behaviour is consistent, a 400 MeV shift is observed between the two CSDA simulations, with MUSIC being the closest to GEANT4, indicating that the CSDA parameters in PUMAS should be further studied. 
PUMAS in backward simulation, in both figures \ref{fig:comparisonPUMAS} and \ref{fig:comparisonMUSIC}, features no events in the lower energy range; this is expected, because a BMC program can only catch the particles which pass through the rock and reach the detector. 
By comparing the behaviour of PUMAS backward mode in figure \ref{fig:comparisonPUMAS}, where the final energy distribution is assumed to be flat within its range, and figure \ref{fig:comparisonMUSIC} where the exact final energy distribution from MUSIC is fed as input, we can see a better agreement with GEANT in the latter case, both in peak value and in shape of the high energy region. Interestingly, a peak below 0.001 GeV can also be seen in the latter case, as a very low but non-zero energy is simulated for the stopping particles.

\subsubsection*{Flux Simulation with PUMAS BMC}
\begin{figure}[h!]
    \centering
    \begin{minipage}{0.45\textwidth}
        \centering
        \includegraphics[width=0.87\textwidth]{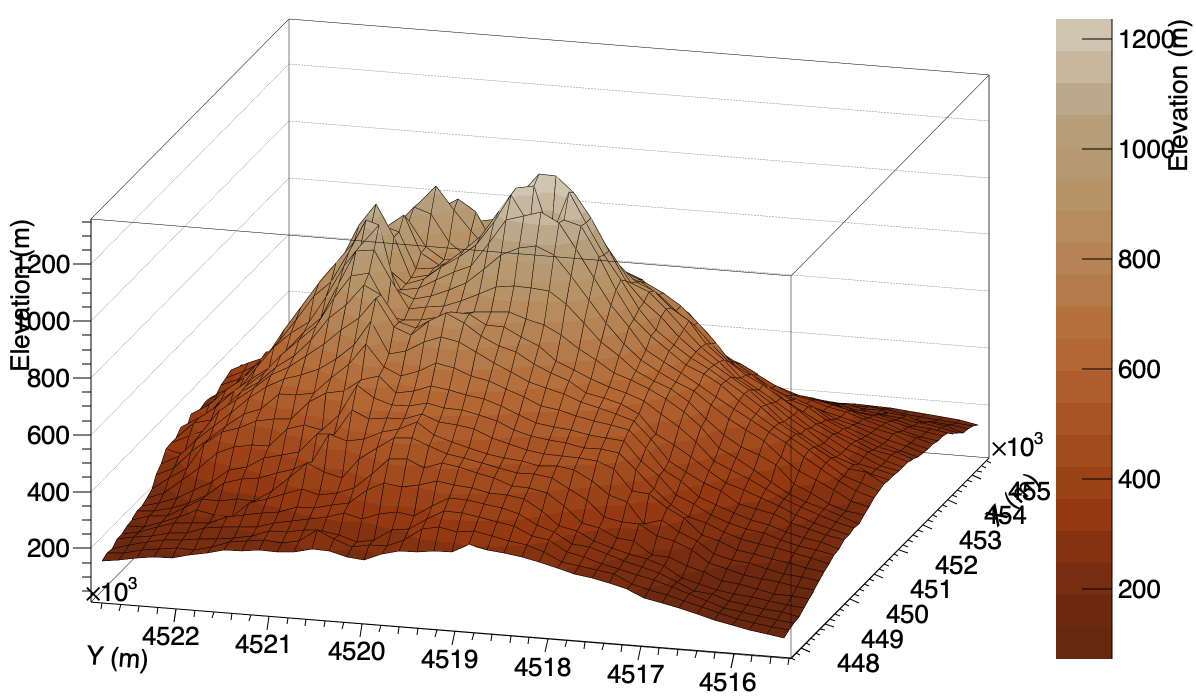}
        \caption{3D visualization of the surrounding area of the Mt. Vesuvius, based on a 5m precision DEM file.}
         \label{fig:dem}
    \end{minipage}\hfill
    \begin{minipage}{0.45\textwidth}
        \centering
        \includegraphics[width=0.9\textwidth]{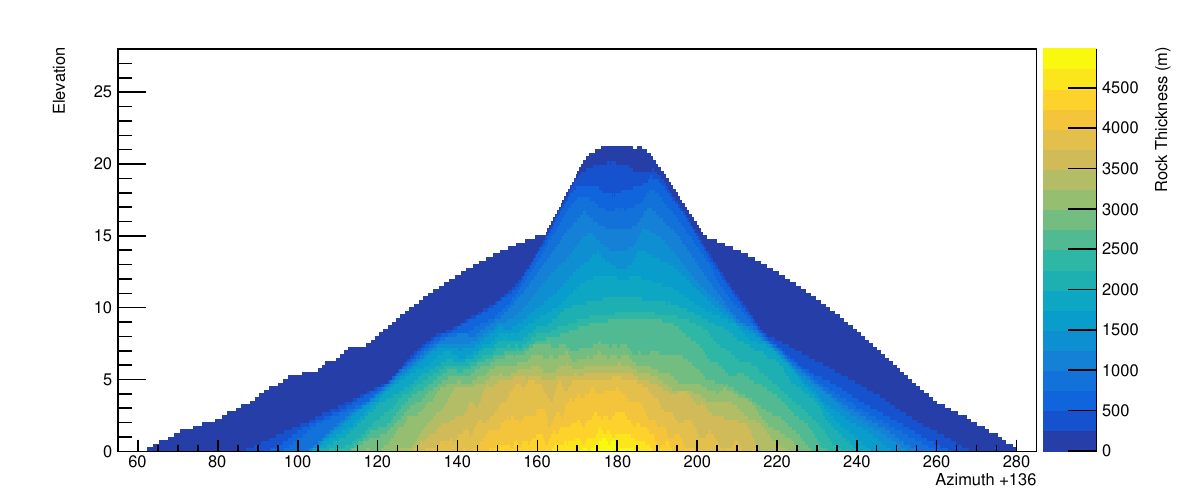}
         \caption{2D thickness map of the Mt. Vesuvius as observed from the location of the MURAVES experiment.}
         \label{fig:thickness}
    \end{minipage}
\end{figure}

In order to compute a transmitted muon flux through an irregularly shaped object such as a volcano, first we need to define the topography. 
To do so, we employ the TURTLE library~\cite{TURTLE}, a utility for the long range transport of MC-generated particles through a predefined topography. Its input is a Digital Elevation Model (DEM), which is a representation of the bare ground on the topographic surface of the Earth excluding trees, buildings, and any other surface object. 
TURTLE reads a DEM file, and user-defined parameters include the observation location, azimuth and elevation angles; the program returns as output a rock thickness map from the observation point. 
In our case, we use a DEM file of the surrounding area of the Mt. Vesuvius with 5~m precision, provided by INGV (National Institute of Geophysics and Volcanology, Naples, Italy) based on data from Ref.~\cite{vilardo2013morphometry}. 
The observation point is obviously set as the location of the MURAVES telescopes.  
A 3D model of Mt. Vesuvius and the surrounding area extracted from this DEM file is visualized in figure \ref{fig:dem}. TURTLE then returns a full scale view of Mt. Vesuvius rock thickness map in azimuth and elevation angles observed at the location of the MURAVES experiment, see figure \ref{fig:thickness}.

\begin{figure}[h]
\centering
\includegraphics[width=10cm]{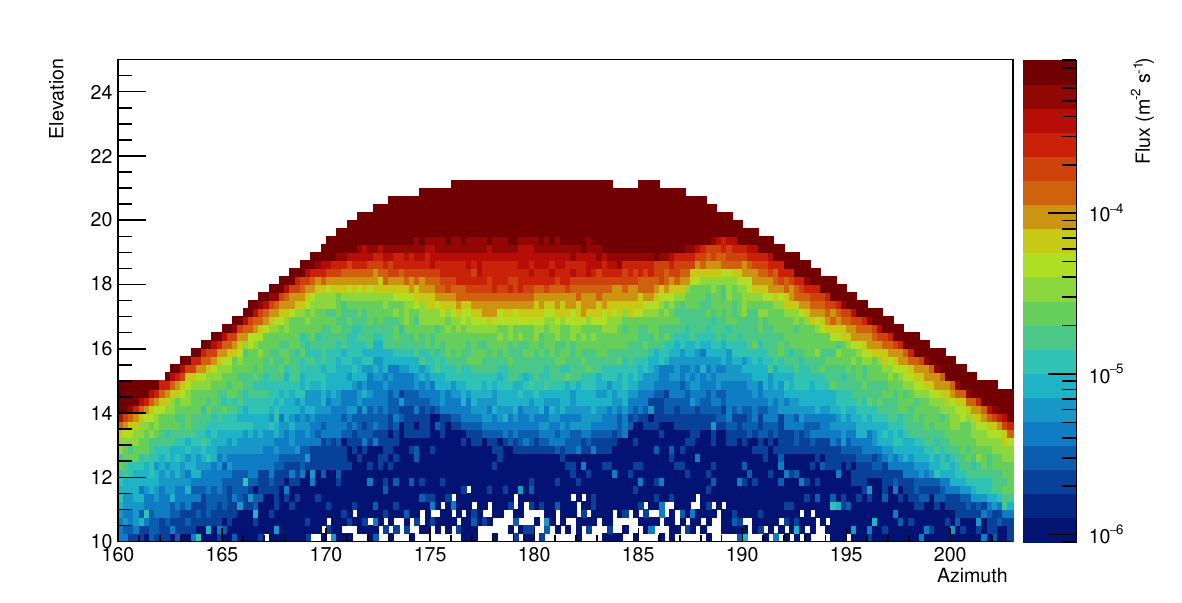}
\caption{Muon flux, predicted with PUMAS, through the Mt. Vesuvius crater as observed from the location of MURAVES, with predefined standard rock density 2.65 $g/cm^3$.}
\label{fig:flux}
\end{figure}

The subsequent phase is to simulate of the muon flux transport through the Mt. Vesuvius using the PUMAS Backward mode. As depicted in Fig. \ref{fig:comparison}, the final energy range as for simulation inputs spans from the minimum to maximum muon energy reaching the MURAVES hodoscope, encompassing the range of 5 MeV to 3000 GeV. Straggled mode is employed in this simulation. A focused examination of the region of interest, namely the crater of Mt. Vesuvius, is presented in Fig. \ref{fig:flux}. The flux rate plot is shown in logarithmic scale, with a color scale emphasizing flux values less than 9E-4. A clear trend is observed, wherein the layers of flux decreasing with an increase in the thickness of Mt. Vesuvius.

It is imperative to note that these results are preliminary; a comprehensive comparison with real data necessitates free-sky simulations and then calculations of transmission rates. On the other hand, work is currently ongoing to set-up also in MUSIC the possibility to read DEM files. Once that becomes available, more refined studies will be performed to decide which between PUMAS (in backward mode) and MUSIC should become the main tool in MURAVES, while the other will definitely still have a place as a cross-check tool. 
In addition to the accuracy of the simulation (quantified by the comparison with the ``golden standard'' provided by GEANT4), the decision will take into account the speed and the simplicity of integration in the full MC chain. 

%Then is to perform the backward transport simulation with PUMAS library, the expected muon flux pass through the Mt. Vesuvius body is obtained. Our interest of region is the summit, we see an agreement between the real data collected in 51 days and the PUMAS simulation.

\section{First results}
\label{sec:first-results}

This section summarizes preliminary work already reported in Refs.~\cite{d2022muraves,MariaelenaThesis}.

\subsection{Data sets}
\label{sec:datasets}

The three telescopes (NERO, ROSSO and BLU) have been deployed between Fall 2019 and Summer 2020, and took data almost continuously since then. 
For each telescope, we separate the datasets according to their orientations (Vesuvius runs and free-sky runs) and to the operating working points of their SiPMs. 

A working point is defined by a target temperature, to which the temperature control system stabilizes the SiPM temperature, and a bias voltage applied to the SiPM which is optimized for the given target temperature. 
The target temperature should ideally be within 5-7 degrees of the environmental temperature in the container, to avoid large power consumption and to stay at safe distance from the dew point and thus avoid damage due to condensation.
A few working points have been defined, and the working point is automatically changed in case of large changes in external temperatures. 
Some performance variations have been observed as a function of the working point, which are also reflected in the trigger rate. Consequently, all results are separately extracted for the different working points, and only combined if statistically consistent. 

\begin{table}[h!]
    \centering
    \begin{tabular}{c|c|c}
       Dataset  &  Vesuvius runs & Free-sky runs \\ \hline
       ROSSO, WP15  & 51 days & 9.5 days \\
       ROSSO, WP20  & 40 days & 14.3 days \\
       NERO, WP15  & 43 days & 10 days \\
       NERO, WP20  & 26 days & 17 days \\
    \end{tabular}
    \caption{Cumulative duration of the Vesuvius and free-sky runs analysed for the first preliminary results.}
    \label{tab:datasets}
\end{table}

\begin{figure}[h!]
    \centering
    \includegraphics[width=0.92\linewidth]{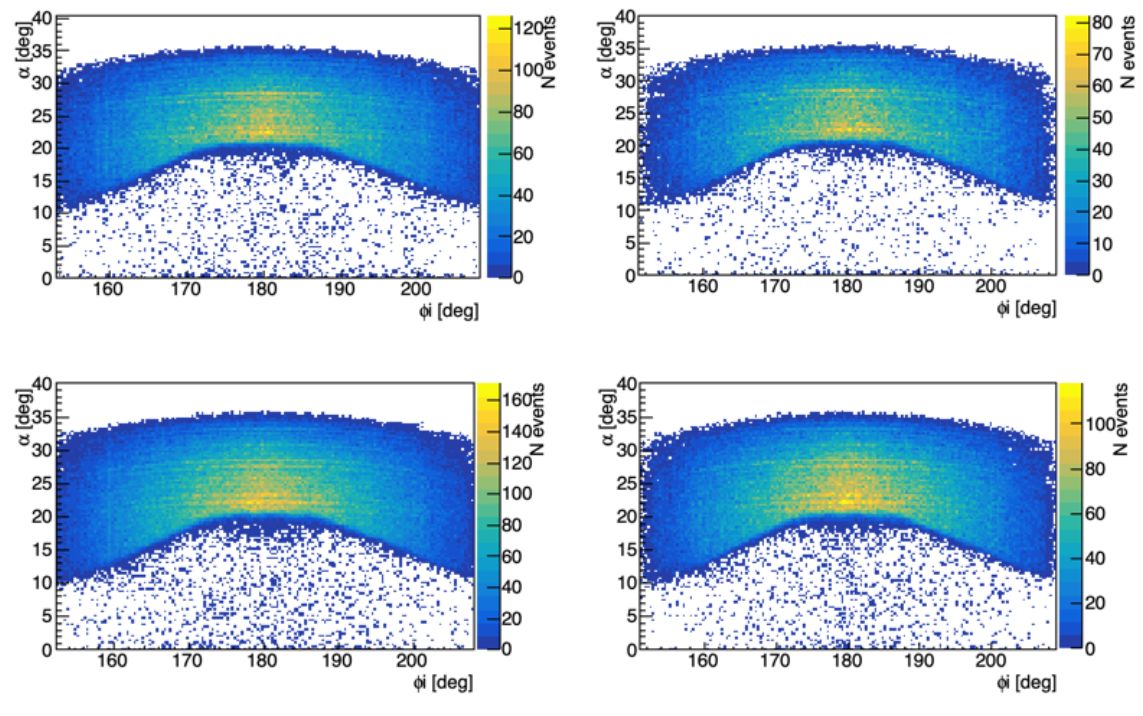}
    \caption{Muon counts as a function of $\alpha$ and $\phi$ for the Vesuvius datasets for NERO (top row) and ROSSO (bottom row) in WP15 (left column) and WP20 (right column).}
    \label{fig:raw-counts-vesuvius}
\end{figure}

\begin{figure}[h!]
    \centering
    \includegraphics[width=0.92\linewidth]{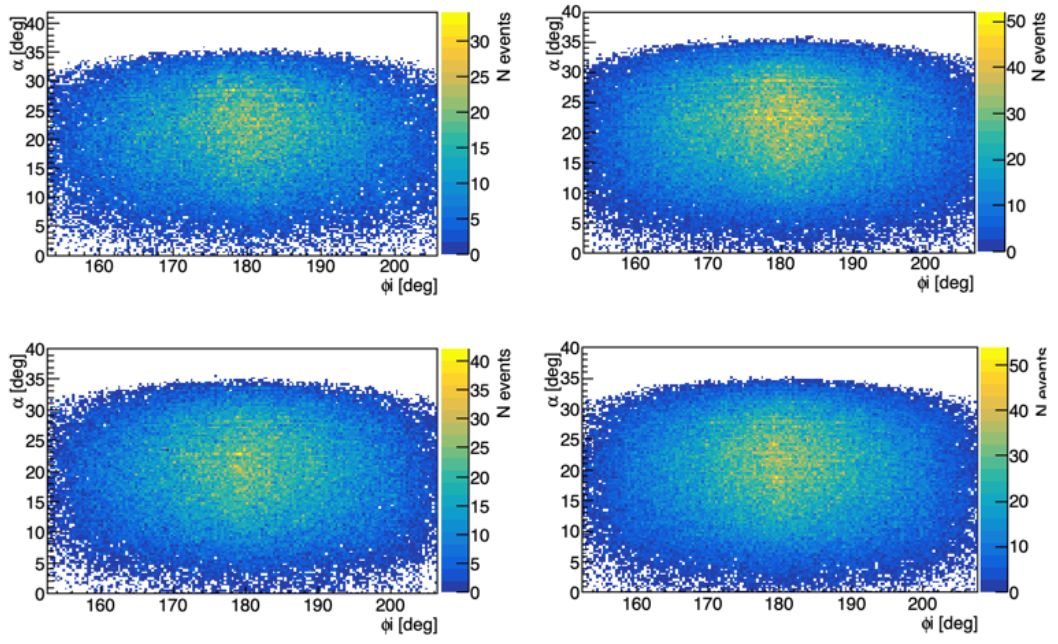}
    \caption{Same as Fig.~\ref{fig:raw-counts-vesuvius} for the free-sky datasets.}
    \label{fig:raw-counts-freesky}
\end{figure}

In this document we only consider well-understood early data from the NERO and ROSSO telescopes, from the first months of MURAVES operations. 
The two main working points, with which most data were collected, correspond to target temperatures of 15 and 20 Celsius degrees, and are denoted as WP15 and WP20, respectively. Other working points (WP5, WP10, WP25) are not considered in this paper as they do not contribute much to the statistics and their performances appear significantly different.
Track quality is quantified by the degree of alignment of the hits from which it is reconstructed, i.e. the $\chi^2$ of the linear fit, normalized by the number of degrees of freedom. And upper cut is applied on the normalized $\chi^2$, but because of the possible differences in performance between different telescopes and different working points, the same $\chi^2$ cut can lead to different track rates in different datasets. To mitigate that, a data-driven procedure has been developed~\cite{MariaelenaThesis} based on the $\chi^2$ distributions observed in $(\Delta\alpha,\Delta\phi)$ control regions in both Vesuvius and free-sky runs. 
Table~\ref{tab:datasets} lists the datasets employed, their duration. %, and the cuts on the normalised $\chi^2$. 

%\begin{table}
%    \centering
%    \begin{tabular}{c|c|c|c}
%       Dataset  &  Vesuvius runs & Free-sky runs & $\chi^2$ cut \\ \hline
%       ROSSO, WP15  & 51 days & 9.5 days & 5.0 \\
%       ROSSO, WP20  & 40 days & 14.3 days & 4.4 \\
%       NERO, WP15  & 43 days & 10 days & 5.1 \\
%       NERO, WP20  & 26 days & 17 days & 5.1 \\
%    \end{tabular}
%    \caption{Cumulative duration of the Vesuvius and free-sky runs analysed for the first preliminary results, and upper cut on the normalised $\chi^2$ applied to the tracks in each dataset selection.}
%    \label{tab:datasets}
%\end{table}

Although the free-sky runs are shorter than the Vesuvius ones, their statistics is much larger, as can be seen by comparing the muon counts in Figs.~\ref{fig:raw-counts-vesuvius} and \ref{fig:raw-counts-freesky} and in general they never contribute significantly to the statistical uncertainty in $T(\alpha,\phi)$ (eq.~\ref{eq:transmission}) for the bins corresponding to the Great Cone.

\subsection{Density projection asymmetries}

With such small statistics of muons surviving the passage through Mt. Vesuvius, it is not possible to produce a very detailed $T(\alpha,\phi)$ map. 
However, we can already do a first measurement of actual interest for volcanology by comparing the muon flux through very large angular bins at different altitudes. 

\begin{figure}[h!]
    \centering
    \includegraphics[width=0.5\linewidth]{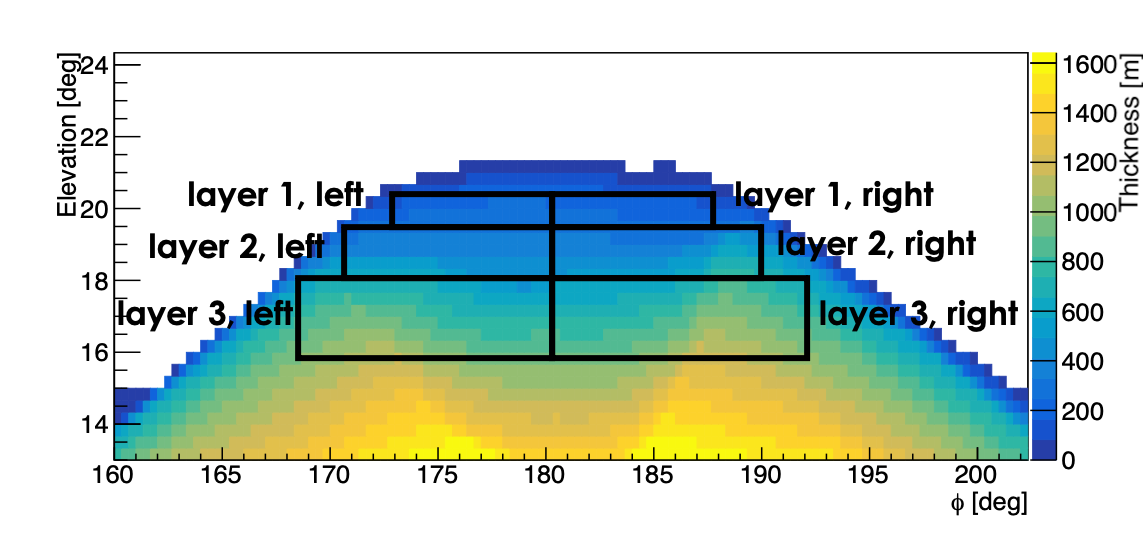}
    \caption{Definition of the angular regions utilized for the first measurement of density projection asymmetries. Reproduced from Ref.~\cite{d2022muraves}.}
    \label{fig:left-right-regions}
\end{figure}

We focus on the summit, i.e. the elevation range $\alpha \ge 16^{\circ}$, which corresponds to $< 1$~km thickness of rock, as shown in Fig.~\ref{fig:left-right-regions}, which also shows the definition of the $(\alpha,\phi)$ regions to be compared. 
To minimize the impact of model assumptions~\footnote{This includes the energy spectrum, for which different Monte Carlo programs give significantly different predictions at low energy (see Sec.~\ref{sec:simulation}), and which is affected by several periodic and aperiodic time-dependent effects that are difficult to model and to correct for, but to which a ratio analysis is insensitive.}, we aim at measuring ratios instead of absolute values of the projected density. 

The muon counts in each region~\footnote{Detailed numerical values can be found in Refs~\cite{MariaelenaThesis,d2022muraves}.} are normalized by thickness of rock traversed, using the DTM provided by INGV~\cite{vilardo2013morphometry}, whose O(m) resolution is not a limiting factor at this level of precision. 
Most modeling uncertainties cancel out in the ratio, and no reliance on simulations is necessary apart from the DTM. 
To be noted that in this measurement the free-sky data are not used in the normalization, although they have been indirectly crucial through their role in the calibration and validation of the detectors. 

%\begin{table}
%    \centering
%    \begin{tabular}{c|c|c|c|c|c|c}
%         & $\Delta\alpha\times\Delta\phi$ & ROSSO WP15 & ROSSO WP20 & NERO WP15 & NERO WP20 & Average \\ \hline
%Layer 1  & $[19.5^{\circ}, 20.5^{\circ}] \times [173^{\circ},187^{\circ}]$ & $1.08^{+0.11}_{-0.09}$ & $1.16^{+0.12}_{-0.10}$ & $1.07^{+0.14}_{-0.11}$ & $1.02^{+0.17}_{-0.13}$ & $1.09^{+0.06}_{-0.05}$\\
%Layer 2  & $[18^{\circ}, 19.5^{\circ}] \times [170^{\circ},190^{\circ}]$ & $0.99^{+0.09}_{-0.08}$ & $0.92^{+0.11}_{-0.09}$ & $0.96^{+0.13}_{-0.10}$ & $0.93^{+0.14}_{-0.11}$ & $0.96^{+0.06}_{-0.05}$\\
%Layer 3  & $[16^{\circ}, 18^{\circ}] \times [168^{\circ},192^{\circ}]$ & $0.87^{+0.09}_{-0.08}$ & $0.92^{+0.09}_{-0.08}$ & $0.94^{+0.11}_{-0.09}$ & $0.91^{+0.14}_{-0.11}$ & $0.90^{+0.05}_{-0.04}$\\
%    \end{tabular}
%    \caption{Preliminary measurements of the right/left opacity asymmetry at three different altitudes, and corresponding statistical uncertainties.} %{\bf Giulio, Gigi, are the angular ranges correct? I tried to guess them from the figure, but it is not easy.}}
%    \label{tab:asymmetry-results}
%\end{table}

The right/left density asymmetry results at three different altitudes are reported in Table~\ref{tab:asymmetry-results}. 
For each layer, the four independent samples agree within one standard deviation ($\sigma$). 
In the last column of the table we report the result of their combination layer by layer, under assumption of statistical independence. 

These numbers suggest ($1.5\sigma$) that the projected density is larger on the right than on the left at high quota, while this relationship inverts at lower quota. 
The analysis of additional data, including the third MURAVES telescope, will be needed in order to prove or disprove these indications.

\begin{table}[h!]
    \centering
    \begin{tabular}{c|c|c|c|c|c}
         & ROSSO WP15 & ROSSO WP20 & NERO WP15 & NERO WP20 & Average \\ \hline
Layer 1  & $1.08^{+0.11}_{-0.09}$ & $1.16^{+0.12}_{-0.10}$ & $1.07^{+0.14}_{-0.11}$ & $1.02^{+0.17}_{-0.13}$ & $1.09^{+0.06}_{-0.05}$\\
Layer 2  & $0.99^{+0.09}_{-0.08}$ & $0.92^{+0.11}_{-0.09}$ & $0.96^{+0.13}_{-0.10}$ & $0.93^{+0.14}_{-0.11}$ & $0.96^{+0.06}_{-0.05}$\\
Layer 3  & $0.87^{+0.09}_{-0.08}$ & $0.92^{+0.09}_{-0.08}$ & $0.94^{+0.11}_{-0.09}$ & $0.91^{+0.14}_{-0.11}$ & $0.90^{+0.05}_{-0.04}$\\
    \end{tabular}
    \caption{Preliminary measurements of the right/left opacity asymmetry at three different altitudes, and corresponding statistical uncertainties.} %{\bf Giulio, Gigi, are the angular ranges correct? I tried to guess them from the figure, but it is not easy.}}
    \label{tab:asymmetry-results}
\end{table}
\section{Conclusions and Prospects}
\label{sec:conclusion}

MURAVES has been taking data smoothly since late 2019. 
Early data, based on few months with two muon telescopes, have been thoroughly validated and used for a right/left density asymmetry measurement at three different altitudes. 

Three years of data are already on disk, with all three telescopes, whose analysis is expected to greatly improve the precision of this and other studies of volcanological interest. 
Higher statistical power, however, imposes a more thorough estimation of the systematic uncertainties. 
This includes detector systematics, which are particularly important for absolute density measurements, and a large variety of modeling systematics whose estimation is only possible via Monte Carlo, such as muon spectrum, time dependent effects, passage through rock, etc.

A full MURAVES simulation chain is being setup. So far, the interface of CRY-MUSIC-GEANT4 chain is setup, meaning that we are able to run the three MC programs sequentially in a single run, without need of saving intermediate data files.  
We are able to obtain accurate transmitted flux predictions from PUMAS in backward mode by defining the Mt. Vesuvius topography with TURTLE and reweighting the spectrum with MUSIC; however, we still lack a complete integration of PUMAS into our MC chain.

\begin{figure}[h]
\centering
\includegraphics[width=14cm]{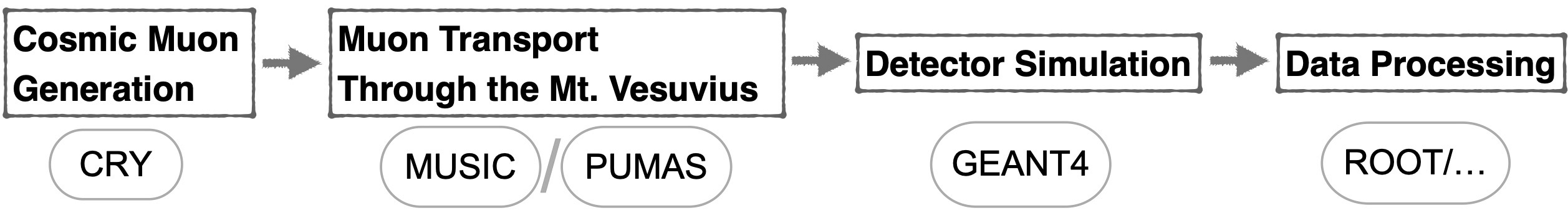}
\caption{Current simulation chain of the MURAVES experiment.}
\label{fig:simulationChain}
\end{figure}

Once a realistic end-to-end Monte Carlo chain will be in place, simulation studies will also allow improvements in track fitting and muon selection, including the introduction of new cuts to reduce the low-momentum component of the spectrum (e.g. by exploiting scattering in the lead wall and time-of-flight from the first to the last layer of the telescope). The global optimization of the analysis chain will have to be based on the achievable resolution, therefore a realistic simulation of the volcano will be crucial, although a trade-off with computation time will be necessary. 
\section*{Acknowledgements}

This work was partially supported by the EU Horizon 2020 Research and Innovation Programme under the Marie Sklodowska-Curie Grant Agreement No. 822185 (``INTENSE'').
Al Moussawi, Basnet and Giammanco also received funding by the Fonds de la Recherche Scientifique - FNRS under Grants No. T.0099.19 and J.0070.21, and by the Federation Wallonie-Bruxelles in the framework of the convention ``FWB-Cellules Europe 2023''. 
Basnet also acknowledges additional funding by the Fund for Research Training in Industry and Agriculture of the Fonds de la Recherche Scientifique - FNRS.

%{\bf INCLUDE ITALIAN FUNDING SOURCES AT THE BEGINNING OF THE PARAGRAPH.}

The authors gratefully acknowledge the precious help by Valentin Niess and Cristina Carloganu; their simulations of Mt. Vesuvius with PUMAS~\cite{PUMAS} were used at the start of the MURAVES project for the choice of the optimal location for the installation of the detectors, and Valentin kindly provided many useful clarifications to the authors of this paper on the correct usage of PUMAS and TURTLE. 
We are also indebted to Pasquale Noli and Nicola Mori for their previous work on the simulation of MU-RAY (precursor of MURAVES)~\cite{GGS}.

\bibliographystyle{unsrt}
\bibliography{refs.bib}

\end{document}